\documentclass[aps,prl,reprint,groupedaddress,showpacs,amsmath,amssymb,twocolumn,floatfix]{revtex4}
\usepackage{txfonts}
\usepackage{graphicx,amssymb,mathrsfs,bm,color,times}
\usepackage{ulem}

\newcommand{\be}{\begin{equation}}

\begin{document}

\title {Phase Space Crystals: A New Way to Create a Quasienergy Band Structure  }

\author{Lingzhen Guo$^{1,2,3}$}
\author{Michael Marthaler$^{1,3}$}
\author{Gerd Sch\"on$^{1,3}$}
\affiliation{$^1$\mbox{Institut f\"ur Theoretische Festk\"orperphysik, Karlsruhe Institute of Technology, 76128 Karlsruhe, Germany}\\
\mbox{$^2$Department of Physics, Beijing Normal University, Beijing 100875, China}\\
$^3$\mbox{ DFG-Center for Functional Nanostructures (CFN),
Karlsruhe Institute of Technology, 76128 Karlsruhe, Germany}\\}

\date{\today}

\begin{abstract}
A novel way to create a band structure of the quasienergy spectrum
for driven systems is proposed based on the discrete symmetry in
phase space. The system, e.g. an ion or ultracold atom trapped in
a potential, shows no spatial periodicity, but it is driven by a
time-dependent field coupling highly nonlinearly to one of its
degrees of freedom (e.g., $\sim q^n$). The band structure in
quasienergy arises as a consequence of the $n$-fold discrete
periodicity in phase space induced by this driving field. We
propose an explicit model to realize such a \textit{phase space
crystal} and analyze its band structure in the frame of a
tight-binding approximation. The phase space crystal opens new
ways to engineer energy band structures, with the added advantage
that its properties can be changed \textit{in situ} by tuning the
driving field's parameters.
\end{abstract}


\maketitle

The high interest in the manipulation of energy band structures,
with the aim to create exotic materials or to tailor their
properties for specific  applications, has opened a research field
of band structure engineering
\cite{Bandengineering1,Bandengineering2}.  The technology relies
on doping \cite{Semiconductor,Semiconductor1} or the application
of external magnetic and electric fields to modify the properties
of
 materials such as semiconductors or graphene
\cite{Graphene,Graphenegap1,Graphenegap2,Graphenegap3}.
Furthermore,
a variety of artificial  periodic structures, such as photonic and
phononic crystals
\cite{Photoniccrystal1,Photoniccrystal2,Photoniccrystal3,Photoniccrystal4,Photoniccrystal6}
or metamaterials
\cite{Metamaterials1,Metamaterials2,Metamaterials3}, are being
investigated to provide band structures optimized for specific
devices.

A system that is driven by a periodic external field shows a
discrete time translation symmetry. In the framework of the
Floquet theory \cite{Floquettheory}
  the concepts of quasienergy and Floquet states \cite{Quasienergy1,Quasienergy2}
were introduced to account for this time periodicity. Normally,
the quasienergy spectrum of a localized system,  e.g., of an ion
trapped in a potential, shows no band structure. But for a
periodically driven crystalline material, as a result of combined
periodicities, the quasienergy spectrum  exhibits a band structure
 \cite{Floqeutquansiband1,Floqeutquansiband2,Floqeutquansiband3,Floqeutquansiband4} in quasimomentum space,
and even a new kind of exotic material, namely, a Floquet
topological insulator \cite{FTI1}, has been proposed.

Here we explore a new discrete symmetry that can be used to create
exotic materials and to manipulate their band structures. The
Hamiltonian of any system depends on two conjugate variables,
momentum and coordinate,  which define the phase space.
 As we will show, it is  possible to create a discrete symmetry in phase space. This leads to specific transformations, which mix
momentum and coordinate, but leave the Hamiltonian unchanged.
 We call such a system a \textit{phase space crystal}. In natural crystals, a periodic potential
 leads to extended states (Bloch states) in real space. The phase space crystal has eigenstates, which are localized in real space
 but are nevertheless energetically so tightly spaced that they
form bands. Since the phase space crystal arises due to driving,
it continuously emits radiation. As a consequence of the band
structure of the quasienergy, the emission spectrum shows
characteristic features, which should be observable experimentally
by methods described in the literature
\cite{EmissionMeasurement1,EmissionMeasurement2}.

{\it Model and RWA}.--- 
As a specific example, we consider a nonlinear oscillator, driven
by an external field coupling nonlinear to the coordinate, with
Hamiltonian
\begin{equation}\label{DDO}
H(t)=\frac{{p}^2}{2m}+\frac{1}{2}m\omega_0^2 q^2 + \frac{\nu}{2}
q^4+2f\cos(\omega_d t){q}^n.
\end{equation}
Here, $\omega_0$ is  the frequency of the oscillator, and
$\omega_d$ is the driving frequency. The nonlinearity is
characterized by the exponent $n$.  If $n=1$, the model
(\ref{DDO}) is the linearly driven Duffing oscillator
\cite{NonlinearOsci}; for $n=2$,  it is a parametrically driven
oscillator \cite{ParametricOsci}. In the present paper we are
interested in the limit of large $n$, say of order $n=10$. There
are various ways to create such high-power coupling. One is based
on so-called "power-law trapping" potentials $V(q)\sim q^n$, which
have been  explored for ultracold atoms
\cite{powerlawtrapping1,powerlawtrapping2,powerlawtrapping3}.
There are reports of static or adiabatically slow changing of the
power-law potential
\cite{powerlawtrappinga,powerlawtrappingb,powerlawtrappingc,powerlawtrappingd,powerlawtrappinge}.
The driving we propose in Eq.(\ref{DDO}) can be realized by making
the power-law trapping potential oscillate with frequency
$\omega_d$. Alternatively, one can create high-power driving terms
by coupling a trapped ion to an external oscillating point charge
or electric dipole. We will further discuss  ways to create $n$th
power driving terms at the end of this paper.

\begin{figure}
\centerline{\includegraphics[height=0.25\textheight]{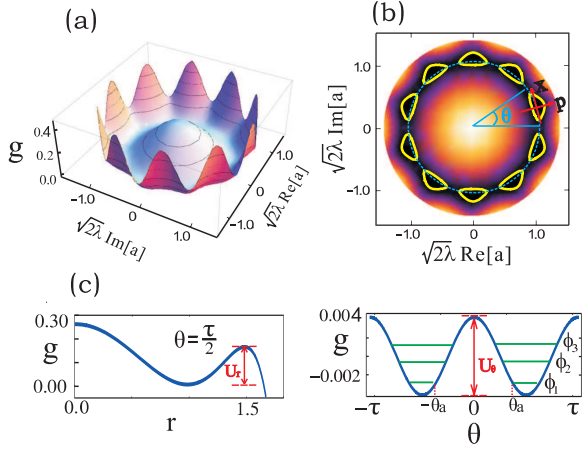}}
\caption{\label{fig1}{ Quasienergy g in phase space.} (a)
$g\propto H_R$  versus Re$[a]$ and Im$[a]$
for power $n=10$ and driving strength $\mu=0.4\mu_c$. For nonzero
driving, the quasienergy is invariant under discrete phase space
rotations $e^{i\theta} \rightarrow e^{i(\theta+\tau)}$ where
$\tau=2\pi/n$. (b) A cut through the bottom of the quasienergy $g$
in (a). There are $n$ stable states (yellow closed curves) and $n$
saddle points (unstable states, between two stable states)
arranged periodically in angular direction. A local coordinate
system $(x,p)$ is defined near the bottom of a stable state. (c)
Quasienergy $g$ versus radius $r$ (left) and angle $\theta$
(right). Stable states are confined by the radial potential
barrier $U_r$ and the angular potential barrier $U_\theta$. For
the latter, we plot two wells between $\theta=-\tau$ and
$\theta=\tau$. The localized states confined in each well (green
lines) are coupled by quantum tunneling.}
\end{figure}

We assume that the driving frequency $\omega_d$ is close to $n$
times $\omega_0$; i.e., the detuning
$\delta\omega\equiv\omega_0-\omega_d/n$ is much smaller than
$\omega_0$. We perform a unitary transformation of the Hamiltonian
$H(t)$ via $\hat{U}=e^{i({\omega_d}/{n}) \hat{a}^\dagger \hat{a}
t}$, where $\hat{a}$ is the annihilation operator of the
oscillator. Dropping  fast oscillating terms, in the spirit of the
rotating wave approximation (RWA), we arrive at the
time-independent Hamiltonian
\begin{equation}\label{H}
\hat{H}_R=\hbar\delta\omega {\hat{a}^\dagger}
\hat{a}+\frac{3\nu\hbar^2}{4m^2\omega_0^2}{\hat{a}^\dagger}
\hat{a} ({\hat{a}^\dagger} \hat{a}+1)+
f\Big(\frac{\hbar}{2m\omega_0}\Big)^{\frac{n}{2}}\Big({\hat{a}^{\dagger
n}}+\hat{a}^n\Big).
\end{equation}
Although RWA is widely used in the
study of driven systems, it is not immediately clear that it is valid
 for highly nonlinear coupling (e.g., $\sim q^n$).
To test it, we performed an exact numerical simulation based on
the full Floquet theory, not relying on the approximation, and
present the results in the Supplemental Material. The conclusion
is that as long as $|\delta\omega|/\omega_0<2\lambda$ (see the
definition of $\lambda$ below) the RWA is well justified.

{\it Discrete symmetry}.--- 
The RWA Hamiltonian Eq.(\ref{H}) displays a new symmetry not
visible in Eq.(\ref{DDO}). To illustrate it, we first make use of
a semiclassical approximation,
 replacing
the operator $\hat{a}$ by a complex number, and plot the resulting
Hamiltonian ${H}_R$ (\ref{H}) in the phase space spanned by
Re$[a]$ and Im$[a]$. The results, seen in Figs.~1(a) and ~1(b),
clearly display the discrete angular periodicity of ${H}_R$. For
the following theoretical analysis, we define a unitary operator
$\hat{T}_\tau=e^{-i\tau {\hat{a}^\dagger} \hat{a}} $ with the
properties $\hat{T}_\tau^\dagger \hat{a} \hat{T}_\tau = \hat{a}
e^{-i\tau}$ and $\hat{T}_\tau^\dagger \hat{a}^n \hat{T}_\tau =
\hat{a}^n e^{-in\tau}$. It is easy to see that the RWA Hamiltonian
is invariant under discrete transformation $T_\tau^\dagger
\hat{H}_R T_\tau = \hat{H}_R$ for $\tau={2\pi}/{n}$.

The discrete angular symmetry suggests introducing the radial and
angular operators $\hat{r}$ and  $\hat{\theta}$ via
$\hat{a}=e^{-i\hat{\theta}}\hat{r}/\sqrt{2\lambda}$ and
$\hat{a}^\dagger=\hat{r}e^{i\hat{\theta}}/\sqrt{2\lambda}$. They
obey the commutation relation
\begin{equation}\label{comm}
[\hat{r}^2,e^{i\hat{\theta}}]=2\lambda e^{i\hat{\theta}}
\end{equation}
 where $\lambda=-3\nu\hbar/(4m^2\omega_0^2\delta\omega)$ is the scaled dimensionless nonlinearity.
Using this definition, we get
$\hat{H}_R=-(\hbar\delta\omega/\lambda) \hat{g}$, with
\begin{equation}\label{quantumg}
\hat{g}=\frac{1}{4}(\hat{r}^2+\lambda-1)^2+\frac{1}{2}\, \mu \,
\left[\left({\hat{r}e^{i\hat{\theta}}}\right)^n+\left({e^{-i\hat{\theta}}\hat{r}}\right)^n
\right].
\end{equation}
The dimensionless driving strength is
$$\mu=-\frac{2\lambda f}{\hbar
\delta \omega}\left( \frac{m\omega_0\delta
\omega}{-3\nu}\right)^{n/2}. $$
For red detuning, $\delta \omega < 0$, considered in the following $\mu > 0$.

{\it Semiclassical analysis}.--- 
We first analyze the properties of the phase space crystal
 in the semiclassical limit $\lambda$$\rightarrow$\ $0$.
For vanishing driving $\mu=0$, the quasienergy $g$ is independent
of the angle $\theta$, which means $g$ is invariant under
continuous phase space rotation. However, for  finite driving
$\mu\neq0$, the quasienergy $g$ is only invariant under discrete
phase space rotations $e^{i\theta} \rightarrow e^{
i(\theta+\tau)}$ with $\tau=2\pi/n$. The periodic arrangement of
atoms in a crystal replaces the continuous translation symmetry by
a discrete one. Similarly, in a phase space crystal the stable
points break the continuous rotation symmetry, and define the
periodicity for the phase space crystal. In Fig.~1(b), the stable
points are the $n$ minima $(r_m,\theta_m)$ of $g$. Between every
two neighboring stable points there is a saddle point
$(r_s,\theta_s)$.

In the vicinity of stable points, the quasienergy $g$ creates
effective potential barriers for angular and radial motion
$U_\theta$ and $U_r$, respectively. Both are shown in Fig.~1(c).
Because of thermal or quantum fluctuations, the states may jump or
tunnel between  neighboring stable points across or through the
angular potential with height $U_\theta\approx 2\mu$. The
tunneling determines the band structure to be discussed below. In
the Supplemental Material, we show that the height of the radial
potential barrier $U_r$ decreases as the driving $\mu$ increases,
up to a critical driving strength $ \mu_c=(1-r^2_c)/(nr^{n-2}_c)$
with $r^2_c=(n-2)/(n-4)$, above which  the stable points
disappear. In the limit of large $n$, we find $
\mu_c\approx{2}/[{\rm e}n(n-2)],$ where ${\rm e}$ is the Euler
constant. In the following, we assume $\mu<\mu_c$ to guarantee the
existence of stable points.

\begin{figure}
\centerline{\includegraphics[scale=0.6]{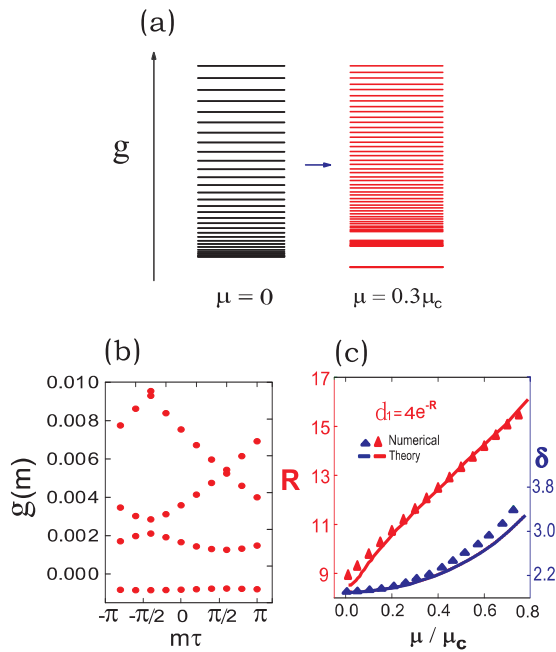}}
\caption{\label{fig2}{Quasienergy band structure.} (a) Quasienergy
spectrum changing from quasicontinuous  in the absence of driving
(left) to a band structure induced by finite driving (right). The
gaps start to open from the bottom of the spectrum. (b)
Quasienergy band structure in the reduced Brillouin zone. Each red
dot represents one quasienergy level. There are $n$ levels in each
band.
 (c) Width of the lowest band $d_1$, and
the asymmetry factor $\delta$ versus driving. Numerical
 (triangles) and approximate (lines) results are compared. The parameters are $\lambda=1/205$,
$n=10$ for all the figures, and for (b) we choose $\mu=0.3\mu_c$.}
\end{figure}

{\it Quasienergy band structure}.--- 
In the quantum regime, $\hat{r}$ and $\hat{\theta}$ no longer
commute. In Fig.~2(a), we show the eigenvalue spectrum of the
quasienergy Hamiltonian obtained from a numerical diagonalization.
In the limit of vanishing driving $\mu\rightarrow 0$, the spectrum
is quasicontinuous whereas for
 $\mu\neq0$ gaps open from the bottom of
the spectrum. According to Bloch's theorem, the eigenstates
$\psi_m(\theta)$ of the quasienergy Hamiltonian
$\hat{g}\psi_m(\theta)=g(m)\psi_m(\theta)$
 have the form $
\psi_m(\theta)=\varphi_m(\theta)e^{-im\theta}$, with a periodic
function $\varphi_m(\theta+\tau)=\varphi_m(\theta)$. Here, the
integer number $m$, which we call a "quasinumber", plays the role
of the quasimomentum $\overset{\rightharpoonup}{\vphantom{a}\smash
k}$ in a crystal. Whereas the quasimomentum
$\overset{\rightharpoonup}{\vphantom{a}\smash k}$ is conjugate to
the coordinate, the quasinumber $m$ is conjugate to the phase
$\theta$.  In Fig.~2(b), we plot the quasienergy band structure in
the reduced Brillouin zone $m\tau\in(-\pi,\pi]$. Here, we relabel
the eigenstates $\psi_m(\theta)$ by $\psi_{lm}(\theta)$, where
$l=1,2,...$ is the label of the bands counted from the bottom. For
finite values of $n$ (in our numerical simulation we chose
$n=10$), the quasienergy band spectrum is discrete. It would
become more continuous in the limit of large $n$.

{\it (i) Band gaps}.--- 
The band structure is characterized by band gaps and bandwidths.
If the driving is  weak, $\mu \ll \mu_c$, only the first gap is
visible. The gaps between higher bands are too narrow to
distinguish them from the level spacings due to finite $n$. In
perturbation expansion, we find for the first gap and bandwidth
$\Delta_1\approx \mu$ and
 $d_1\approx{\lambda^2n^2}/{4}-{\mu}/{2}+{\mu^2}/({2\lambda^2n^2})$, respectively.
I.e., the gap $\Delta_1$ increases linearly with the driving,
whereas the bandwidth $d_1$ decreases with driving. For stronger
driving, the spectrum of the $l$th band is approximately
\begin{eqnarray}\label{spectruminoneband}
g_l(m)=E_l-2|J_l|\cos(m\tau+\delta\tau),
\end{eqnarray}
centered around  $E_l$ and with bandwidth $d_l=4|J_l|$. The result
shows a surprising asymmetry. From the plot of the quasienergy in
Fig.~1(b), we would have expected a degeneracy $g(m)=g(-m)$, since
clockwise and anticlockwise motion should be equivalent, as in the
case of orbital motion.
 However, in the present case, the two degrees of freedom of phase space Im$[a]$ and Re$[a]$ do not
 commute, and as a result the quasienergy structure is asymmetric.
The degree of asymmetry is characterized in
Eq.~(\ref{spectruminoneband}) by the asymmetry factor $\delta$.

In the case of sufficiently strong driving, several levels are
localized in each stable point, as shown by Fig.~1(c) (right
figure). The band structure can be explained by a tight-binding
model: the gaps are opened by level spacings of localized states
at the same stable point, whereas the bandwidth is determined by
quantum tunneling between nearest neighbors. At the bottom of each
stable point, to lowest order, the localized Hamiltonian can be
approximated by a harmonic form with
 effective frequency $\omega_e=\left\{\mu n^2
r_m^{n-2}[3r_m^2-1-n(n-1)\mu r_m^{n-2}]\right\}^{1/2}$ (see the
Supplemental Material). Since $r_m\approx1$, the localized quantum
level spacing is $\lambda \omega_e\approx n\lambda\sqrt{2\mu}$.
The level spacing corresponds to the distance between two centers
of adjacent bands.  The anharmonicity leads to higher-order
corrections to the level spacings, for levels close to the bottom
proportional to $-l\lambda^2$, where $l$ is the label of the band.
This negative correction means that higher level spacings decrease
linearly.
The tight-binding approximation is valid for a
$\mu>\lambda\omega_e/2$, where the angular potential barrier
$U_\theta\approx2\mu$ is high enough to confine at least one
quantum level in each stable point.

{\it (ii) Asymmetry factor}.--- 
The most unusual feature of the band structure
(\ref{spectruminoneband}) is the asymmetry characterized by the
factor $\delta$. It results from the following property of the
operator $\hat{r}^2$: in $\theta$ representation, one could
conclude that the operator $\hat{r}^2$ with form
$-i2\lambda{\partial}/{\partial \theta}$ satisfies  the
commutation relation (\ref{comm}) exactly. However, in this case
the eigenvalues of $\hat{r}^2$ could be negative, which would be
unphysical. We, therefore, define a local coordinate system
$(x,p)$ measured from the bottom of a stable point as shown in
Fig.~1(b). In the limit of large $n$, we have local operators
$\hat{x}\approx\bar{r}(\hat{\theta}-\tau/2)$ and
$\hat{p}=\hat{r}-\bar{r}$,
where $\bar{r}$ is the average radius. Their commutation relation
is $[\hat{p},\hat{x}]=i\lambda$. Thus, in ``$x$ representation" ,
we have
$\hat{p}=i\lambda({\partial}/{\partial x}),$
and $\hat{r}=\bar{r}+\hat{p}=\bar{r}+i\lambda({\partial}/{\partial
x})$.  Dropping the $\lambda^2$ term, we get
$\hat{r}^2\approx\bar{r}^2+2i\lambda\bar{r}({\partial}/{\partial
x})$. As a result, the first term of quantum quasienergy
Hamiltonian (\ref{quantumg}) becomes
$[2i\lambda\bar{r}({\partial}/{\partial
x})+\bar{r}^2+\lambda-1]^2/4$, which indeed distinguishes
anticlockwise and clockwise direction since
$\bar{r}^2+\lambda-1\neq0$ in general.  In addition, the driving
term in the Hamiltonian (\ref{quantumg}) introduces some asymmetry
by changing the average radius $\bar{r}$.

We can explicitly calculate the asymmetry factor $\delta$ in the
frame of the tight-binding model.  The relation between the Bloch
eigenstate $\psi_{lm}(\theta)$ and the localized state in each
stable point $\phi_l(\theta)$, as indicated in Fig.~1(c), is given
by
$\psi_{lm}(\theta)=1/\sqrt{n}\sum_{q=0}^{n-1}e^{imq\tau}{\hat{T}_\tau^q}\phi_l(\theta)$.
Only the nearest neighbor coupling
$J_l=-\int[\hat{T}_\tau\phi_l(\theta)]^*\hat{g}\phi_l(\theta)d\theta$,
is important. From $\hat{T}_\tau\phi_l(\theta) \approx
e^{-i\tau\bar{r}^2/2\lambda}\phi_l(\theta+\tau)$, it follows to be
$J_l=-e^{i\tau\bar{r}^2/2\lambda}\int\phi_l^*(\theta+\tau)\hat{g}\phi_l(\theta)d\theta=|J_l|e^{i\tau\bar{r}^2/2\lambda}$.
The corresponding quasienergy spectrum of the $l$th band then is
$g_l(m)=\int_0^{2\pi}\psi^*_{lm}(\theta)\hat{g}\psi_{lm}(\theta)d\theta\approx
E_l-J_le^{im\tau}-J_l^*e^{-im\tau}=E_l-2|J_l|\cos(m\tau+\bar{r}^2\tau/2\lambda)$.
Hence the asymmetry factor is $\delta=\bar{r}^2/2\lambda \
(\mathrm{mod}\ n)$. A similar phase shift for the tunneling
amplitude has been found for the special case of the parametric
oscillator ($n=2$) in Ref. \onlinecite{ParametricTunneling}. For
the bottom band, the average radius is
$\bar{r}_1=1-\lambda/2+\sum_{k=1}^{\infty}\bar{c}_{2k} \mu^{2k}$
with average coefficient $\bar{c}_{2k}$ given in the Supplemental
Material. To get the average radius of next higher levels, we use
the quantization condition in phase space
$(\bar{r}^2_{l+1}-\bar{r}^2_l)\tau/2=\pi \lambda$.

In Fig.~2(c), we show the dependence of the asymmetry factor
$\delta$ on the driving strength $\mu$, obtained in both the
tight-binding calculation described above and from a numerical
simulation. The asymmetry arises from the phase of the complex
tunneling parameter $J_l=|J_l|e^{i\tau\bar{r}^2/2\lambda}$. The
phase factor $\tau\bar{r}^2/2\lambda$ called Peierls phase
\cite{peierlsphase1,peierlsphase2} has also been discussed as a
possibility to realize artificial gauge fields
\cite{peierlsphase2,artificialgauge1} for ultracold atoms. For
optical lattices, there are already some proposals to create a
controlled Peierls phase by synthesizing a one-dimensional
effective Zeeman lattice \cite{artificialgauge2} or shaking the
lattice \cite{peierlsphase1}. In the present case, the complex
tunneling parameter $J_l$ naturally arises in the plane of the
phase space.

{\it (iii) Bandwidths}.--- 
The $l$th bandwidth is $d_l=4|J_l|$. To calculate the amplitude of
the coupling $|J_l|$, we use the double-well potential model, as
shown by the right plot in Fig.~1(c). For the analysis of quantum
tunneling, the property of quasienergy near the saddle point
($r_s,\theta_s$) is important. We move the local coordinate system
$(x,p)$ defined above to the saddle point ($r_s,\theta_s=0$). Now
the local coordinates are given by $x\approx r_s\theta$ and $p=
r-r_s$. To  second order, the Hamiltonian near the saddle point
can be approximated by
\begin{equation}
 g\approx  \frac{1}{2}m_s \omega^2_s p^2+\mu r^n_s \cos
 (n\theta)+\frac{(r^2_s-1)^2}{4},
\end{equation}
where $m_s\omega^2_s=\partial^2 g/\partial r^2|_{(r_s,\theta_s)}$.
Given an energy level $E_l$, one can write $|p|$ as a function of
$\theta$ and calculate the amplitude of the coupling
\begin{equation}\label{J}
 |J_l|=\frac{\lambda \omega_e}{2\pi}\exp[-\frac{r_s}{\lambda}\int_{-\theta_a}^{\theta_a}|p|d\theta].
\end{equation}
Here, $\theta_a$ is the turning point that is given by
$\theta_a=1/n\cos^{-1}[(E_l-{(r^2_m-1)^2}/{4})/\mu r^n_m]$. The
 integral in the exponent of Eq.~(\ref{J}) is given by
\begin{equation}
 \int_{-\theta_a}^{\theta_a}|p|d\theta=\frac{4}{n}\Big\{\frac{2}{m_s \omega^2_s}\Big[\mu r^n_s
 +\frac{(r^2_s-1)^2}{4}-E_l\Big]\Big\}^{1/2}
 E(\phi|k),
\end{equation}
where $E(\phi|k)=\int_0^\phi\{1-k^2\sin^2\theta\}^{1/2}d\theta$ is
the elliptic integral of the second kind with  parameters
$\phi=n\theta_a/2$ and $k=\left\{2\mu r^n_s/[\mu r^n_s
+{(r^2_s-1)^2}/{4}-E_l]\right\}^{1/2}$. In Fig.~2(c), we compare
our approximate result for the first bandwidth $d_1$ versus
driving $\mu$ to  numerical results. In the tight-binding regime
they agree well with each other.

{\it Emission spectrum}.--- 
The above calculation of quasienergy band structure does not
account for a dissipative environment. It renders the time
evolution of phase space crystal nonunitary and induces
transitions between quasienergy states
\cite{MarkOldDDO,ParametricActivated}. For a driven quantum
system, even at base temperature $T=0$ many quasienergy states can
be excited and transitions between them will contribute to the
emission spectrum
\cite{ParametricSpectrum,QuantumHeatingMeasured}. The spectral
density of the photons emitted by the driven resonator
\cite{DDOspectrum} follows from $S(\omega) = 2 \, \mathrm{Re}
\int_0^{\infty} d  t \, \langle a^{\dagger} ( t) a \rangle_{\rm
st} e^{-i \omega t}$.

\begin{figure}[tp]
\begin{center}
\begin{tabular}{ll}
    \includegraphics[scale=0.5]{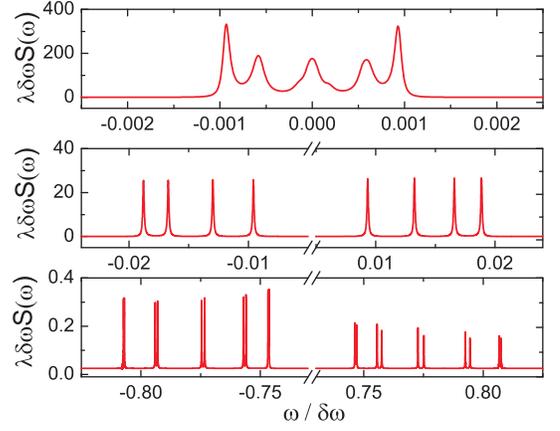} \\
\end{tabular}
\caption{\footnotesize{Emission spectrum:  The top, middle and
bottom figures are the emission spectrum for the first band, the
second band and the interband respectively. The parameters:
$\lambda=1/205$, temperature $\mathrm{T}=0.5\hbar\omega_0/k_B$,
driving $\mu=0.4\mu_c$, the damping $\kappa=10^{-4}\lambda$.}}
\label{spectrum}
\end{center}
\end{figure}

 To calculate
the correlation function $C(t)=\langle a^{\dagger} (t) a
\rangle_{\rm st}$, we need a master equation that also accounts
for the dissipative evolution caused by thermal and quantum
fluctuations. He have checked that a Lindblad-type master equation
\cite{ParametricSpectrum,MarkOldDDO,LindbladME1,LindbladME2,LindbladME3}
is sufficient for the present situation,
\begin{equation}\label{ME}
\frac{\partial \rho}{\partial \tau}=-\frac{i}{\lambda}[g,{\rho}]
 + \kappa(1+\bar{n}){\cal D}[a]{\rho} +  \kappa \bar{n}{\cal
 D}[a^{\dagger}]{\rho}=\cal{L}\rho.
\end{equation}
The dimensionless time $\tau=t \, \delta\omega$ is scaled by the
detuning. The Lindblad superoperator is defined through ${\cal
D}[A]{\rho}\equiv A{\rho}A^{\dagger}
 -(A^{\dagger}A{\rho}+{\rho}A^{\dagger}A )/2 $ where $\bar{n}=(e^{\hbar\omega_0/k_BT}-1)^{-1}$
is the Bose distribution and $\kappa$ is the dimensionless damping
scaled by the detuning. We make use of the quantum regression
theorem to calculate the correlation function, i.e.,
$C(\tau)=\mathrm{Tr}[a^{\dagger} ( \tau) a \rho_{\rm
st}]=\mathrm{Tr}[a^{\dagger} e^{{\cal{L}}\tau } (a \rho_{\rm
st})]$. The spectral density $S(\omega)$ is the Fourier
transformation of the correlation function $C(\tau)$. We choose
our parameters to confine two localized states in each well; i.e.,
we truncate  our numerical simulation at $2n$ levels.

The total spectrum can be divided into three parts, as shown in
Fig.~3. The top and middle figures represent intraband transitions
of the first and second band, respectively. The bottom figure
corresponds to interband transition between the first and second
bands. The positive and negative frequencies in the emission
spectrum correspond to absorption of energy from and emission of
energy to the driving field, respectively. The widths of the peaks
in emission spectrum are proportional to the damping $\kappa$. The
quasienergy band structure can be directly detected by analyzing
the spectrum of emitted photons in the laboratory. It should be
noticed, however, that the above emission spectrum is obtained in
the rotating frame with frequency $\omega_d/n$. Hence, a value of
$\omega$ in this spectrum represents a photon with frequency
$\omega+\omega_d/n$ in the laboratory frame.

{\it Discussion}.--- 
The phase space crystal is a general consequence of a discrete
rotation symmetry in phase space and is not restricted to the
model presented in detail above. More generally it can be found
for Hamiltonians such as $H(t)={{p}^2}/{2m}+m\omega_0^2 q^2/2+V(q)
+ f(q)\cos(\omega_d t)$. For cold atoms, the nonlinear driving can
be created by using power-law trapping methods. For trapped ions,
it can be caused by an oscillating point charge coupling to the
charged ion via Coulomb interaction, leading to the expansion
$f(q)\propto 1/(1-q)=\sum^{\infty}_{k=0}q^k$. In the parameter
range where RWA is valid (i.e., for
$|\delta\omega|/\omega_0<2\lambda$ as derived in the Supplemental
Material), in combination with the resonance condition $ \omega_d
\approx n\omega_0$, the driving term will automatically pick up
terms $a^n$ and $a^{\dagger n}$ from $q^n$, or terms $a^\dagger
a^{n+1}$ and $a^{\dagger n+1} a$ from $q^{n+2}$, etc. All these
RWA terms remain invariant under discrete phase space rotation
$e^{i\theta}\rightarrow e^{i(\theta+2\pi/n)}$. In the model
analyzed above, we further assumed a nonlinear static potential
$V(q)=\nu q^4/2$. Also, this can be chosen to be more general. If
$V(q)$ is an even function of coordinate $q$, the RWA terms with
equal numbers of $a^\dagger$ and $a$ will contribute to the phase
space crystal.

In the solid-state band theory, the spectrum ultimately becomes
continuous due to the large number of atoms. For the phase space
crystal, a continuous quasienergy spectrum would emerge in the
limit of large $n$. Compared to conventional artificial materials,
such as photonic crystals, the energy band structure of phase
space crystals can be changed {\it in situ} by tuning the driving
field's parameters. By changing the coupling power $n$, one can
even change the lattice constant $\tau=2\pi/n$ of the phase space
crystal.  The new symmetry introduces the quasinumber space. The
concept of quasinumber space may bring a new perspective to modify
properties of materials.

{\it Acknowledgements}.--- 
 We acknowledge helpful discussions
with P. Kotetes, J. Michelsen and V. Peano.  L. Guo
acknowledges the support from the China Scholarship Council.

\section*{Supplemental Material}

\subsection{Justification for RWA}

To derive Eq.(2) in the main text we adapted the well-known
rotating wave approximation (RWA). The main results in this paper
were derived within this approximation. In this section we perform
an exact numerical simulation based on the full Floquet theory and
calculate the quasienergy spectrum. We find the condition for the
validity of RWA but also show numerical results beyond the RWA
regime. We also discuss the role of non-RWA terms in the phase
space crystal.

We start from the original time-dependent Hamiltonian in
laboratory frame
\begin{equation}\label{DDOsm}
\hat{H}(q,t)=\frac{{p}^2}{2m}+\frac{1}{2}m\omega_0^2 q^2 +
\frac{\nu}{2} q^4+2f\cos(\omega_d t){q}^n.
\end{equation}
We transform to the rotating frame via
$\hat{U}=e^{i({\omega_d}/{n}) \hat{a}^\dagger \hat{a} t}$ and keep
all the terms
\begin{eqnarray}
\hat{H}_{RF}(t)
=\hat{U}\hat{H}(t)\hat{U}^\dagger+i\hbar\dot{\hat{U}}\hat{U}^\dagger=\hat{H}_{RWA}+\hat{H}_{non-RWA}.
\end{eqnarray}
The first part is the RWA Hamiltonian
$\hat{H}_{RWA}=-(\hbar\delta\omega/\lambda) \hat{g}$ given by
Eq.(2) and Eq.(4) in the main text. The non-RWA part Hamiltonian
$\hat{H}_{non-RWA}$ has the following form
\begin{eqnarray}
\hat{H}_{non-RWA}&=&
\frac{\nu\hbar^2}{4m^2\omega_0^2}(2{\hat{a}^\dagger}\hat{a}-1)\hat{a}^{\dagger2}e^{i{2\omega_dt}/{n}}+
\frac{\nu\hbar^2}{8m^2\omega_0^2}\hat{a}^{\dagger4}e^{i{4\omega_dt}/{n}}\nonumber\\
&&+f\Big(\frac{\hbar}{2m\omega_0}\Big)^{\frac{n}{2}}\Big[\Big({\hat{a}^{\dagger
}e^{i{2\omega_dt}/{n}}}+\hat{a}\Big)^n-\hat{a}^n\Big]+\textit{h.c.}.
\end{eqnarray}
It is time-dependent, but the total  Hamiltonian $\hat{H}_{RF}(t)$
is periodic, $\hat{H}_{RF}(t)=\hat{H}_{RF}(t+T)$, with period
$T=2\pi/(2\omega_d/n)=n\pi/\omega_d$. In the framework of Floquet
theory, the solution of Schr\"odinger equation
$i\hbar\frac{\partial}{\partial
t}|\Psi(t)\rangle=\hat{H}_{RF}(t)|\Psi(t)\rangle$ has the form
$|\Psi(t)\rangle=e^{-i\epsilon t/\hbar}|\psi(t)\rangle$, where
$|\psi(t)\rangle$ is the \textit{Floquet state} satisfying
$|\psi(t)\rangle=|\psi(t+T)\rangle$\cite{Floquettheory}. Then the
Schr\"odinger equation becomes
\begin{equation}\label{SEsm}
 \mathscr{H}(t)|\psi(t)\rangle=\epsilon|\psi(t)\rangle.
\end{equation}
Here,
$\mathscr{H}(t)=\hat{H}_{RF}(t)-i\hbar\frac{\partial}{\partial t}$
is the \textit{Floquet Hamiltonian} and $\epsilon$ is termed the
\textit{quasienergy}.

In order to calculate the quasienergy spectrum, we need to
diagonalize the Floquet Hamiltonain $\mathscr{H}(t)$. Because of
the periodicity $\mathscr{H}(t)=\mathscr{H}(t+T)$, it is
convenient to introduce a composite Hilbert space
$\mathscr{R}\otimes\mathscr{T}$, where $\mathscr{R}$ is the
spatial space with the time-independent basis $|\phi_m\rangle
\in\mathscr{R}$, which are determined by the eigenstates of RWA
Hamiltonian $\hat{g}$
 \begin{equation}
 \hat{g}|\phi_m\rangle=g_m|\phi_m\rangle,
\end{equation}
while $\mathscr{T}$ is the space of functions with time
periodicity $T$. We can choose the time-dependent Fourier vectors
$\langle t|q\rangle=\exp(iq\frac{2\omega_d}{n}t)$ with
$q=0,\pm1,\pm2,\ldots,$ as the orthonormal basis of space
$\mathscr{T}$. We denote the eigenstates and eigenvalues of the
Floquet Hamiltonian $\mathscr{H}(t)$ by $|\psi_{m,q}(t)\rangle$
and $\epsilon_{m,q}$,
\begin{equation}\label{floqueteigenfunction}
\mathscr{H}(t)|\psi_{m,q}(t)\rangle=\epsilon_{m,q}|\psi_{m,q}(t)\rangle.
\end{equation}
 Under the
RWA, it is easy to see that
$|\psi_{m,q}(t)\rangle=e^{-iq\frac{2\omega_d}{n}t}|\phi_m\rangle$
and $\epsilon_{m,q}=-(\hbar\delta\omega/\lambda)
g_{m}+{2q\hbar\omega_d}/{n}$, i.e., for $q\neq0$, the quasienergy
spectrum is shifted by ${2q\hbar\omega_d}/{n}$.

Eq.(\ref{SEsm}) has infinitely many equivalent solutions. This is
because the Floquet state $|\psi(t)\rangle$  is allowed to be
time-dependent. After a time-dependent gauge transformation, the
state $e^{-iq\frac{2\omega_d}{n}t}|\psi(t)\rangle$ is still a
solution of Eq.(\ref{SEsm}), with corresponding quasienergy
shifted \cite{Floquettheory} by ${2q\hbar\omega_d}/{n}$, that is,
$$\mathscr{H}(t)\Big(e^{-iq\frac{2\omega_d}{n}t}|\psi(t)\rangle\Big)=(\epsilon+{2q\hbar\omega_d}/{n})\Big(e^{-iq\frac{2\omega_d}{n}t}|\psi(t)\rangle\Big),$$
where $q=0,\pm1,\pm2,\ldots,$. Thus we can map all the states with
$q\neq0$ to the state with $q=0$. The full form of this eigenstate
$|\psi_{m,0}(t)\rangle$ with time-dependent Fourier expansion is
\begin{equation}\label{c0sm}
|\psi_{m,0}(t)\rangle=C^0_m|\phi^0_m\rangle+\sum_{m',q\neq0}C_{m'q}e^{iq\frac{2\omega_d}{n}t}|\phi_{m'}\rangle.
\end{equation}
The corresponding quasienergy will  also be modified, that is,
$\epsilon_{m,0}=-(\hbar\delta\omega/\lambda) g_{m}+\Delta_m$.
Here, the renormalized state $|\phi^0_m\rangle$ is in general a
superposition of spatial basis states $|\phi_m\rangle$. The first
term on the right side of Eq.(\ref{c0sm}) is a time-independent
term which represents the RWA part while the second term of
Eq.(\ref{c0sm}) represents the contribution from non-RWA
Hamiltonian $\hat{H}_{non-RWA}$. Thus quantity $P_0=|C^0_m|^2$ is
the probability for the RWA part of the full state.

Both the RWA probability $P_0$ and the quasienergy
$\epsilon_{m,0}$ are functions of the detuning $\delta\omega$. As
long as it is small enough, $|\delta\omega|/\omega_0\ll1$, the RWA
works well, which means $P_0\approx1$ and
$\epsilon_{m,0}\approx-(\hbar\delta\omega/\lambda) g_{m}$.
However, for stronger detuning  more and more higher order
oscillating modes should be included as indicated by the sum in
the second term of Eq.(\ref{c0sm}). By exact numerical simulation,
we can calculate the relationship between $P_0$ and detuning as
shown in Fig.~4a). We see that there is a critical point
$|\delta\omega_c|$ for each $\lambda$ (see the definition of
$\lambda$ in the main text). When the absolute value of detuning
is smaller than the critical value, i.e.,
$|\delta\omega|<|\delta\omega_c|$, the RWA is well justified. The
critical value $|\delta\omega_c|$ depends on the parameter
$\lambda$. We can use the following simple method to estimate its
value. Since the RWA Hamiltonian can be written as
$\hat{H}_{RWA}=-(\hbar\delta\omega/\lambda) \hat{g}$, where
$\hat{g}$ is a scaled dimensionless quantity, the prefactor
$-(\hbar\delta\omega/\lambda)\equiv\hbar|\delta\omega|/\lambda$
(we assume a red detuning, $\delta \omega < 0$) represents the
energy scale of RAW Hamiltonian. In the non-RWA Hamiltonian
$\hat{H}_{non-RWA}$, the lowest oscillating frequency is
$2\omega_d/n$. Thus, the valid regime for the RWA can be estimated
by the following condition
\begin{equation}\label{RWAcondition}
\hbar|\delta\omega|/\lambda<2\hbar\omega_d/n \Longrightarrow
|\delta\omega|/(\omega_d/n)\approx|\delta\omega|/\omega_0<2\lambda
.
\end{equation}
The above condition means the critical point is
$|\delta\omega_c|=2\lambda\omega_0$. On the plot of Fig.~1a), we
indicate the critical points calculated from condition
(\ref{RWAcondition}) for different $\lambda$'s by vertical dashed
lines. They agree with numerical results very well.

In the main text, we show that the quasienergy band structure
comes from the discrete angular rotation symmetry in phase space.
This symmetry is a property of the RWA Hamiltonian. The non-RWA
Hamiltonian $\hat{H}_{non-RWA}$, however, does not have this
discrete symmetry. In fact, the existence of $\hat{H}_{non-RWA}$
will deteriorate the discrete angular rotation symmetry, thus
modify the band structure of the quasienergy spectrum. As shown in
Fig.~1b), large detuning tends to reduce the bandgap and broaden
the bandwidth. But for a large region beyond the RWA regime, the
band structure is very robust to the detuning (the bandwidth stays
much smaller than the bandgap). We may consider non-RWA terms to
behave like ``disorder" in a phase space crystal. Further
investigation of these effects will be independent future work. We
also notice that the bandgap shows some peaks with changing
detuning, and the bandwidth shows dips accordingly. For an
explanation of these peaks beyond RWA see the work by V. Peano et
al. \cite{BeyondRWA}.

\begin{figure}
\centerline{\includegraphics[scale=1.0]{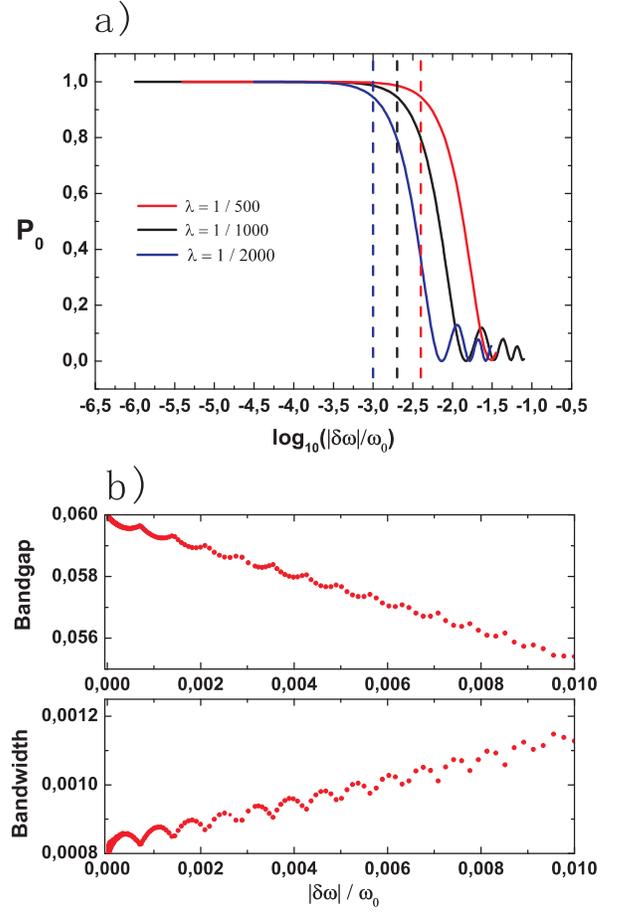}}
\caption{\label{fig1}{The role of strong detuning.} a) The
probability of the RWA part, $P_0$, of the full Floquet state (see
Eq.(\ref{c0sm})) versus detuning for different values of
$\lambda$. Each level exhibits almost the same behavior against
detuning for a fixed $\lambda$. The three colored vertical dashed
lines indicate the critical values according to condition
(\ref{RWAcondition}). b) The relationships between the bandgap
(top) and the bandwidth (bottom), in units of $\hbar\omega_d/n$,
as functions of detuning for $\lambda=1/2000$. Here, we plot the
banddap and bandwidth of the first band. In fact, every band shows
a similar behavior v.s. detuning. For stronger detuning more
higher order oscillating modes should be included in the expansion
of Eq.(\ref{c0sm}). }
\end{figure}

\subsection{ Stability Analysis}

We calculate the extrema of quasienergy in phase space by standard
stability analysis. These extrema are classified into stable
points and saddle points (unstable points). In semiclassical limit
$\lambda\rightarrow 0$, the quasienergy is
\begin{eqnarray}\label{g}
g=\frac{1}{4}(r^2-1)^2+\mu r^n\cos(n\theta).
\end{eqnarray}
 The extrema $(r_e,\theta_e)$ in angular
 and radial direction can be
obtained from
\begin{eqnarray}\label{extreme1a}
\frac{\partial g}{\partial \theta}
\big|_{\theta=\theta_e,r=r_e}&=& -\mu n r_e^n \sin(n\theta_e)=0\\
\label{extreme1b} \frac{\partial g}{\partial r
}\big|_{\theta=\theta_e,r=r_e}&=& r_e[(r_e^2-1)+\mu
nr_e^{n-2}\cos(n\theta_e)] =0 \, .
\end{eqnarray}
The two equations have a trivial solution $r_e=0$. In addition
nontrivial solutions of  the angular dependence can be obtained
from Eq.(\ref{extreme1a}), namely $\theta_e=l\tau/2$ with $l=0,\pm
1,\pm 2,...,\pm (n-1), n$, where $\tau=2\pi/n$ is defined as
\textit{lattice constant} of the phase space crystal. The
corresponding radial extrema can be obtained from
Eq.(\ref{extreme1b})
\begin{eqnarray}\label{re}
r_e=1+\sum_{k=1}^{\infty}c_k(\theta_e) \mu^k.
\end{eqnarray}
Here, the series expansion coefficient $c_k(\theta)$ are given by
\begin{eqnarray}\label{ck}
c_k(\theta)=\frac{(-1)^k({n}/{2})^k\cos^k(n\theta)[k(n-2)-1]!!}{k![k(n-4)+1]!!}.
\end{eqnarray}

The stability of these extrema $(r_e,\theta_e)$ is determined by
the second derivatives of $g$. If $({\partial^2 g}/{\partial
\theta}^2)\times({\partial^2 g}/{\partial r}^2)
\big|_{r=r_e,\theta=\theta_e}>0$, the extrema are stable,
otherwise unstable. From
\begin{eqnarray}
\frac{\partial^2 g}{\partial \theta^2}
\big|_{\theta=\theta_e,r=r_e}= -\mu n^2 r_e^n
\sin(n\theta_e)\big|_{\theta=\theta_e}=-\mu n^2 r_e^n \sin(l\pi).
\end{eqnarray}
we see that odd integers of $l$ give $({\partial^2 g}/{\partial
\theta}^2) \big|_{\theta=\theta_e}>0$, while even integers of $l$
give $({\partial^2 g}/{\partial \theta}^2)
\big|_{\theta=\theta_e}<0$. The second derivative with repect to
the radius $r$ is
\begin{eqnarray}\label{r2derivative}
\frac{\partial^2 g}{\partial r^2 }\big|_{r=r_e,\theta=\theta_e}=
3r^2_e-1+n(n-1)\mu \cos(n\theta_e).
\end{eqnarray}
For weak driving $\mu\ll1$, since the radial extreme is
$r_e\approx 1$, the above value is positive. From this we conclude
that
 the angular extrema at $\theta_m=l\tau/2$   with
$l$ odd integers between $-n$ and $n$ are stable points (minima),
while the angular extrema at $\theta_s=l\tau/2$ with $l$ even
integers between $-n$ and $n$ are unstable saddle points.

As the driving strength $\mu$ increases  the condition
(\ref{r2derivative}) can reduce to zero, which means the
nontrivial solutions of Eq.(\ref{extreme1b}) disappear. The
critical driving $\mu_c$ is determined by
\begin{eqnarray}\label{criticaldriving}
\frac{\partial g}{\partial r }\big|_{\mu=\mu_c}= 0,\ \ \ \ \
\mathrm{and}\ \ \ \ \  \frac{\partial^2 g}{\partial r^2
}\big|_{\mu=\mu_c}= 0 .
\end{eqnarray}
Solving the above two equations,  we get
$\mu_c=(1-r^2_c)/(nr^{n-2}_c)$ with $r^2_c=(n-2)/(n-4) $. In the
 limit of large $n$, the critical driving $
\mu_c\approx{2}/[en(n-2)], $ where $e$ is the Euler constant.

\subsection{Local Hamiltonian }

In this section, we give a perturbative form of the Hamiltonian
close to the bottom of the stable points $(r_m,\theta_m)$. The
eigenvalues and eigenstates of the local Hamiltonian are needed
for the tight-binding calculation. We first write the local
Hamiltonian in a harmonic approximation
\begin{eqnarray}\label{localisedg}
g_{local}&\approx&  \frac{1}{2}\frac{\partial^2 g}{r^2_m\partial
\theta^2
}\big|_{(r_m,\theta_m)}(r_m\theta-r_m\theta_m)^2\nonumber\\
&&+\frac{1}{2}\frac{\partial^2 g}{\partial r^2
}\big|_{(r_m,\theta_m)}(r-r_m)^2+ g(r_m,\theta_m)\nonumber\\
&=&\frac{p^2}{2m_e}
+\frac{1}{2}m_e\omega^2_ex^2+\frac{(r^2_m-1)^2}{4}-\mu r_m^n .
\end{eqnarray}
Here, we have defined the coordinate $x=r-r_m$ and momentum
$p=r_m(\theta-\theta_m)$ near the stable point. The effective mass
$m_e$ and effective frequency $\omega_e$ are given by $
m_e=r_m^{2}({\partial^2 g}/{\partial \theta^2})^{-1}$ and $
\omega_e=\sqrt{m^{-1}_e{\partial^2 g}/{\partial r^2}}$
respectively, with explicit formulars $ m_e=(\mu n^2
r_m^{n-2})^{-1},\ \ \omega_e=\sqrt{\mu n^2
r_m^{n-2}[3r_m^2-1-n(n-1)\mu r_m^{n-2}]}.$

The anharmonicity gives higher order corrections to the localized
Hamiltonian. We transform the original $\hat{g}$ to a local
Hamiltonian $\hat{g}_{local}$ at the stable point $(r_m,\theta_m)$
by three steps. Firstly, we change the orientation using the phase
space rotation operator $\hat{T}_{\theta_m}=e^{-i\theta_m
{\hat{a}^\dagger} \hat{a}}$, resulting in a properly orientated
Hamiltonian $\hat{T}_{\theta_m}\hat{g}\hat{T}^\dagger_{\theta_m}$.
Secondly, we move
$\hat{T}_{\theta_m}\hat{g}\hat{T}^\dagger_{\theta_m}$ to the
position of stable point using the displacement operator
$\hat{D}_\alpha=e^{\alpha a^\dagger-\alpha^* a}$, resulting in a
Hamiltonian sitting at the bottom of stable point
$\hat{D}_\alpha\hat{T}_{\theta_m}\hat{g}\hat{T}^\dagger_{\theta_m}\hat{D}^\dagger_\alpha$.
Finally, we squeeze the Hamiltonian to fit the stable point by
using the squeezing operator $\hat{S}_\xi=e^{[\xi^* a^2-\xi
(a^\dagger)^2]/2}$, resulting in the needed local Hamiltonian
$\hat{g}_{local}=\hat{S}_\xi\hat{D}_\alpha\hat{T}_{\theta_m}\hat{g}\hat{T}^\dagger_{\theta_m}\hat{D}^\dagger_\alpha\hat{S}^\dagger_\xi$.

The displacement operator $\hat{D}_\alpha$ has the property
$D_\alpha^\dagger a D_\alpha=a+\alpha$, while the squeezing
operator $\hat{S}_\xi$ satisfies
$S_\xi^\dagger a S_\xi=va+ua^\dagger$,
where $v=\cosh|\xi|,\ \ u=-\xi/|\xi| \sinh|\xi|$ are the squeezing
parameters. Starting from the following original form of $\hat{g}$
\begin{eqnarray}
\hat{g}=\frac{1}{4}(2\lambda\hat{a}^\dagger\hat{a}+\lambda-1)^2+\frac{1}{2}\mu(2\lambda)^{\frac{n}{2}}
\Big({\hat{a}^{\dagger n}}+\hat{a}^n\Big),\\
\nonumber
\end{eqnarray}
and  choosing the parameters  $\alpha=-r_e/\sqrt{2\lambda}$,
$v=(\sqrt{m_e \omega_e}+1/\sqrt{m_e \omega_e})/2$, and
$u=(\sqrt{m_e \omega_e}-1/\sqrt{m_e \omega_e})/2$ we get the local
Hamiltonian
\begin{eqnarray}\label{localisedghighorder}
\hat{g}_{local}&=&\hat{S}_\xi\hat{D}_\alpha\hat{T}_{\theta_m}\hat{g}\hat{T}^\dagger_{\theta_m}\hat{D}^\dagger_\alpha\hat{S}^\dagger_\xi\nonumber\\
&=&
\lambda\omega_e(\hat{a}^\dagger\hat{a}+\frac{1}{2})+\frac{(r^2_m-1)^2}{4}-\mu
r_m^n +\lambda^{3/2}\Delta\hat{g}+o(\lambda^2).\nonumber\\
\end{eqnarray}
The term $\Delta \hat{g} = (v-u)r_e((v \hat{a}^\dagger - u
\hat{a})(v \hat{a} - u
\hat{a}^\dagger)+1/2)(\hat{a}+\hat{a}^\dagger)/\sqrt{2}-n(n-1)(n-2)\mu
r_e^{n-3}(v \hat{a}^\dagger-v \hat{a})^3/(3\sqrt{2})+c.c.$ is
treated in perturbation theory. In this way we can determine the
local quantum level to order $\lambda^2$.


\subsection{Average radius $\bar{\textit{\textbf{r}}}$}

In order to get the asymmetry factor $\delta$, we need to
calculate the average radius $\bar{r}$.  In the limit of large
$n$, we define a local coordinate system $(x,p)$ near the bottom
of stable points with corresponding operators defined by
$\hat{x}=\bar{r}(\hat{\theta}-\tau/2)$ and
$\hat{p}=\hat{r}-\bar{r}$,
where $\bar{r}$ is the average radius. They satisfy the
commutation relation $[\hat{p},\hat{x}]=i\lambda$. In
``$x$-representation" or ``$\theta$-representation", we have
$\hat{p}=i\lambda\frac{\partial}{\partial
x}=i\frac{\lambda}{\bar{r}}\frac{\partial}{\partial \theta}.$
Then we have $\hat{r}^2/(2\lambda)
=(\bar{r}+\hat{p})^2/(2\lambda)=(\bar{r}^2+2i\lambda\frac{\partial}{\partial
\theta}-\lambda^2\frac{\partial^2}{\partial
\theta^2})/(2\lambda)$. Neglecting terms of order $\lambda^2$, we
get an important relationship $\hat{r}^2/(2\lambda)\approx
\bar{r}^2/(2\lambda)+i\frac{\partial}{\partial \theta}$.

The average radius of the bottom band, $\bar{r}_1$, can be
estimated by averaging $r_e(\theta)$, given by Eq.(\ref{re}), over
the angular direction
$\bar{r}_1=\frac{1}{2\pi}\int_0^{2\pi}r_e(\theta)d\theta.$
 Since
$
\overline{\cos^k(n\theta)}=\frac{1}{2\pi}\int_0^{2\pi}\cos^k(n\theta)d\theta={(k-1)!!}/{k!!}$
for even integer $k$ and $ \overline{\cos^k(n\theta)}= 0$ for odd
integer $k$, we have from Eq.(\ref{ck})
\begin{equation}
 \bar{c}_{2k}=(-\frac{n}{2})^{2k}\frac{(2k-1)!![2k(n-2)-1]!!}{(2k)!!(2k)![2k(n-4)+1]!!}.
\end{equation}
The average radius of the bottom band is given by
$\bar{r}_1=1+\sum_{k=1}^{\infty}\bar{c}_{2k} \mu^{2k}$. This
result is obtained based on the semiclassical quasienergy
(\ref{g}). Considering quantum correction, the final result is
$\bar{r}_1=1-\lambda/2+\sum_{k=1}^{\infty}\bar{c}_{2k} \mu^{2k}$.
This approximation is justified by our numerical simulation.


\begin{thebibliography}{27}
\expandafter\ifx\csname
natexlab\endcsname\relax\def\natexlab#1{#1}\fi
\expandafter\ifx\csname bibnamefont\endcsname\relax
  \def\bibnamefont#1{#1}\fi
\expandafter\ifx\csname bibfnamefont\endcsname\relax
  \def\bibfnamefont#1{#1}\fi
\expandafter\ifx\csname citenamefont\endcsname\relax
  \def\citenamefont#1{#1}\fi
\expandafter\ifx\csname url\endcsname\relax
  \def\url#1{\texttt{#1}}\fi
\expandafter\ifx\csname
urlprefix\endcsname\relax\def\urlprefix{URL} \fi
\providecommand{\bibinfo}[2]{#2}
\providecommand{\eprint}[2][]{\url{#2}}



\bibitem{Bandengineering1}
J. Zhang \textit{et al.} Nature Commun. \textbf{2}, 574 (2011).

\bibitem{Bandengineering2}
 J. R. Beresford, \textit{Band Structure Engineering for Electron
Tunneling Devices}, Columbia University, 1990.



\bibitem{Semiconductor}
P. M. Koenraad and  M. E. Flatt$\mathrm{\acute{e}}$, Nature
Materials \textbf{10}, 91-100 (2011).

\bibitem{Semiconductor1}
 Y. Nishi and R. Doering,  \textit{Handbook of Semiconductor Manufacturing Technology}.
Marcel Dekker Inc., 2000.




\bibitem{Graphene}
 F. Guinea, M. I. Katsnelson and A. K. Geim, Nature Physics \textbf{6}, 30-33 (2010).

\bibitem{Graphenegap1}
K. S. Novoselov \textit{et al}., Science \textbf{306}, 666-669
(2004).


\bibitem{Graphenegap2}
F. Yavari \textit{et al}., Small \textbf{6}, 2535-2538 (2010).


\bibitem{Graphenegap3}
 E. V. Castro \textit{et al}.,  Phys.
Rev. Lett. \textbf{99}, 216802 (2007).



\bibitem{Photoniccrystal1}
J. D. Joannopoulos, P. R. Villeneuve and S. Fan, Nature
\textbf{386}, 143-149 (1997).

\bibitem{Photoniccrystal2}
 E. Yablonovitch, Phys. Rev. Lett. \textbf{58},
2059-2062 (1987); E. Yablonovitch \textit{et al}.,  Phys. Rev.
Lett. \textbf{67}, 2295-2298 (1991).

\bibitem{Photoniccrystal3}
S. John, Phys. Rev. Lett. \textbf{58}, 2486-2489 (1987).

\bibitem{Photoniccrystal4}
Z. V. Vardeny, A. Nahata and A. Agrawal, Nature Photonics
\textbf{7}, 177-187 (2013).

\bibitem{Photoniccrystal6}
E. L. Thomas,  T. Gorishnyy and  M. Maldovan, Nature Materials
\textbf{5}, 773-774 (2006).




\bibitem{Metamaterials1}
M. Choi \textit{et al.}, Nature \textbf{470}, 369-373 (2011).

\bibitem{Metamaterials2}
J. T. Shen, P. B. Catrysse and S. Fan, Phys. Rev. Lett.
\textbf{94}, 197401 (2005).

\bibitem{Metamaterials3}
N. Fang \textit{et al}., Nature Materials \textbf{5}, 452-456
(2006).



\bibitem{Floquettheory}
M. Grifoni and P. H\"anggi, Phys. Rep. \textbf{304}, 229 (1998).

\bibitem{Quasienergy1}
J. H. Shirley, Phys. Rev. \textbf{138}, 4B, 979 (1965)

\bibitem{Quasienergy2}
Y. B. Zeldovitch,  Sov. Phys. JETP \textbf{ 24} (1967) 1006 [Zh.
Eksp. Teor. Fiz. \textbf{51} (1966) 1492].




\bibitem{Floqeutquansiband1}
Z. Gu \textit{et al.},  Phys. Rev. Lett. \textbf{107}, 216601
(2011).

\bibitem{Floqeutquansiband2}
 B. H. Wu \textit{et al.},  Appl. Phys. Lett.
\textbf{100}, 203106 (2012).

\bibitem{Floqeutquansiband3}
 E. S. Morell and Luis E. F. Foa Torres, Phys. Rev. B \textbf{86}, 125449 (2012); H. L. Calvo
\textit{et al.},  Appl. Phys. Lett. \textbf{98}, 232103 (2011); H.
 L. Calvo \textit{et al.},  Appl. Phys. Lett. \textbf{101}, 253506
(2012).

\bibitem{Floqeutquansiband4}
A. G\'omez-Le\'on and G. Platero, Phys. Rev. Lett. \textbf{110},
200403 (2013).

\bibitem{FTI1}
N. H. Lindner \textit{et al.}, Nature Physics \textbf{7}, 490-495
(2011).


\bibitem{EmissionMeasurement1}
C. Stambaugh and H. B. Chan, Phys. Rev. Lett. \textbf{97}, 110602
(2006).

\bibitem{EmissionMeasurement2}
H. B. Chan and C. Stambaugh, Phys. Rev. B {\bf 73}, 224301(2006).


\bibitem{NonlinearOsci}
M. I. Dykman,  Zh. Eksp. Teor. Fiz. {\bf 68}, 2082 (1975).

\bibitem{ParametricOsci}
M. I. Dykman, Phys. Rev. E {\bf 57} , 5202 (1998).

\bibitem{ParametricTunneling}
M. Marthaler and M. I. Dykman, Phys. Rev. A {\bf 76}, 010102(R)
(2007).

\bibitem{powerlawtrapping1}
P. W. H. Pinkse \textit{et al}., Phys. Rev. Lett. \textbf{78},
990-993 (1997)

\bibitem{powerlawtrapping2}
D. M. Stamper-Kurn \textit{et al}., Phys. Rev. Lett. \textbf{81},
2194-2197 (1998)

\bibitem{powerlawtrapping3}
Franco Dalfovo \textit{et al.}, Rev. Mod. Phys. \textbf{71},
463-512 (1999)

\bibitem{powerlawtrappinga}
V. Bagnato, D. E. Pritchard, and D. Kleppner, Phys. Rev. A
\textbf{35}, 4354 (1987).

\bibitem{powerlawtrappingb}
 A. Jaouadi, \textit{et al.}, Phys. Rev. A \textbf{82}, 023613 (2010); A. Jaouadi,
et al., Phys. Rev. A \textbf{ 83}, 023616 (2011).

\bibitem{powerlawtrappingc}
Guozhen Su, Jincan Chen, and Lixuan Chen, Physics Letters A
\textbf{315} (2003) 109-19.

\bibitem{powerlawtrappingd}
Shukuan Cai,\textit{et al.}, Physica A \textbf{387} (2008)
4814-820.

\bibitem{powerlawtrappinge}
 Berna G\"ulveren, Solid State Sciences \textbf{14} (2012),94-99.

\bibitem{peierlsphase1}
J. Struck \textit{et al.}, Phys. Rev. Lett. \textbf{108}, 225304
(2012)

\bibitem{peierlsphase2}
N. Goldman \textit{et al.},arXiv:1308.6533

\bibitem{artificialgauge1}
Jean Dalibard \textit{et al.}, Rev. Mod. Phys.
\textbf{83},1523-1543 (2011)

\bibitem{artificialgauge2}
K. Jim\'{e}nez-Garc\'{i}a, \textit{et al.}, Phys. Rev. Lett.
\textbf{108}, 225303 (2012)

\bibitem{MarkOldDDO}
 M. I. Dykman and V. N. Smelyanskiy, Zh. Eksp. Teor. Fiz. {\bf
94}, 61 (1988).

\bibitem{ParametricActivated}
 M. Marthaler and  M. I. Dykman, Phys. Rev. A
\textbf{73}, 042108 (2006).

\bibitem{ParametricSpectrum}
M. I. Dykman,  M. Marthaler and V. Peano, Phys. Rev. A {\bf 83},
052115 (2011).

\bibitem{DDOspectrum}
 S. Andr\'e,  L. Guo, V. Peano, M. Marthaler and G. Sch\"on,
Phys. Rev. A \textbf{85}, 053825 (2012).

\bibitem{QuantumHeatingMeasured}
F. R. Ong \textit{et al}.,  Phys. Rev. Lett. \textbf{110}, 047001
(2013).

\bibitem{LindbladME1}
L. Guo \textit{et al}., Phys. Rev. E {\bf 84}, 011144 (2011)

\bibitem{LindbladME2}
S. Diehl \textit{et al}.,  Phys. Rev. Lett. \textbf{105}, 015702
(2010).

\bibitem{LindbladME3}
M. I. Dykman, Phys. Rev. E {\bf 75}, 011101 (2007).

\bibitem{BeyondRWA}
V. Peano, M. Marthaler, and M. Dykman, Phys. Rev. Lett.
\textbf{109}, 090401 (2012).



\end{thebibliography}
\end{document}